Real-Time Observation of Self-Interstitial Reactions on an Atomically Smooth Silicon Surface

S. Kosolobov,[a,*] G. Nazarikov,[a] S. Sitnikov,[b] I. Pshenichnyuk,[a] L. Fedina,[b] A. Latyshev[b,c]

[a]Skolkovo Institute of Science and Technology, 100 Novaya Street, Skolkovo, Moscow Region, 143025 Russia; email: sergey.kosolobov@gmail.com

[b]Institute of Semiconductor Physics SB RAS, Acad. Lavrent'eva St. 13, Novosibirsk 630090, Russia

[c]Novosibirsk State University, Pirogova St. 2, Novosibirsk 630090, Russia

**Abstract**

Self-diffusion and impurity diffusion both play crucial roles in the fabrication of semiconductor nanostructures with high surface-to-volume ratios. However, experimental studies of bulk-surface reactions of point defects in semiconductors are strongly hampered by extremely low concentrations and difficulties in the visualization of single point defects in the crystal lattice. Herein we report the first real-time experimental observation of the self-interstitial reactions on a large atomically smooth silicon surface. We show that non-equilibrium self-interstitials generated in silicon bulk during gold diffusion in the temperature range 860-1000°C are annihilated at the (111) surface, producing the net mass flux of silicon from the bulk to the surface. The kinetics of the two-dimensional islands formed by self-interstitials are dominated by the reactions at the atomic step edges. The activation energy for the interaction of self-interstitials with the surface and energy barrier for gold penetration into the silicon bulk through the surface are estimated. These results demonstrating that surface morphology can be profoundly affected by surface-bulk reactions should have important implications for the development of nanoscale fabrication techniques.

1. Introduction

Diffusion in solids is a fundamental material process that is mediated by native point defects. In metals, these defects are known as vacancies and play a prominent role in the kinetics of bulk and surface atomic reactions [1]. For semiconductors, particularly silicon, the situation is more complex: two types of defects, vacancies and self-interstitials, are involved in self-, dopant, and impurity diffusion [2–7]. At nanoscale, due to the increasing surface-to-volume ratio, the atomic reactions of point defects at the surfaces and interfaces become predominant and may significantly modify the properties of fabricated nanostructures [8–11].

While the properties of point defects in bulk semiconductors have been extensively studied and are well understood, little is actually known about the point defect formation, reactions, and basic mechanisms of the diffusion in the vicinity of free surfaces. At the heart of the problem lies the need for direct visualization of the single point defect diffusion in crystal bulk, which is still lacking, especially at high temperatures, where diffusion is fast compared to the imaging rate of modern atomic-scale characterization techniques such as aberration-corrected high-resolution transmission electron microscopy (HRTEM). In addition, the extremely low concentrations of point defects in silicon make their observation quite difficult. On the other hand, the direct visualization of single atoms, vacancies, and their clusters on crystal surfaces has become routine with the development of surface-sensitive techniques such as scanning tunneling (STM), atomic force (AFM), low-energy (LEEM), and reflection electron microscopy (REM). However, atomic steps that are an inherent part of all crystalline surfaces, acting as sinks for adsorbed atoms and vacancies, hamper direct experimental studies of point defect properties.

This study used high-temperature (860-1000°C) submonolayer gold deposition onto an atomically smooth large size (up to 100 μm in diameter) step-free Si(111) surface to study the interaction of the native point defects (self-interstitials and vacancies) with the surface boundary, considering the finite efficiency for the penetration, annihilation, and creation of point defects at the surface. Since it is well established that gold atoms diffuse in silicon bulk via indirect diffusion mechanisms that involve intrinsic point defects, gold deposition is used to increase the concentration of point defects. Employing ultrahigh vacuum reflection electron microscopy and numerical simulations, we demonstrate that silicon self-interstitials formed in a subsurface area during gold diffusion into the silicon bulk diffuse and are annihilated at the crystal surface. By analyzing the kinetics of the surface morphology transformations, we show that the annihilation of the self-interstitials occurs predominantly near the atomic step edges rather than over the entire surface.

2. Materials and methods

As it is extremely complicated to prepare perfectly flat surfaces by cutting the crystal along the needed crystallography orientation, additional efforts are required to prepare step-free areas with large singular terraces. In our experiments, silicon samples were prepared from an n-type phosphorous-doped dislocation-free FZ-grown silicon wafer with a resistivity of 0.3 Ω cm and 0.1° miscut from the (111) plane. Square craters with sides of 500 x 500 μm$^2$ and depths of approximately 1 μm were formed on the substrates via optical lithography. The wafer was cut into rectangular pieces (1.3 × 8 mm$^2$) containing single craters. Then the sample was thermally annealed in an ultrahigh vacuum at 1300°C for several minutes to remove the oxide layer and possible contamination from the preparation technique. After the cleaning procedure, the sample temperature was decreased to 900°C and large step-free areas formed at the bottom of the craters as has been described previously [12].

The experimental investigations of the silicon surface morphology transformations were performed via ultrahigh vacuum reflection electron microscopy (UHV REM) technique, which visualizes the silicon surface morphology in a wide temperature range enabling the in situ characterization of the atomic processes governing the surface morphology formation [13]. Due to the technical design and construction features, REM magnification in the direction parallel to the electron beam incidence is much smaller compared to that in the perpendicular plane. The REM images are therefore compressed in the direction of the electron beam incidence with a factor of 30-50, as indicated by the scale bars. All of the REM images were recorded using a TVIPS FastScan-F114 CCD camera with a frame rate of 25 fps. Subsequently, the video sequences were analyzed frame by frame to investigate the surface dynamics with a time resolution of 0.04 s.

Submonolayer gold deposition was carried out in UHV REM by thermal evaporation from a specially designed small-sized evaporator equipped with a resistivity-heated tungsten crucible and a mechanical shutter. The deposition rate was calibrated in situ by analyzing the diffraction patterns and REM images during gold deposition on the Si(111) surface at 500$^0$C. At this temperature, gold adsorption initiates the formation of an Si(111)-5x2-Au superstructure leading to the appearance of additional reflections in the RHEED patterns. The amount of gold corresponding to the surface fully covered by 5 x 2-Au domains was used as a reference point and assumed to be 0.43 ± 0.14 monolayers (ML), where 1 ML of gold corresponds to 7.8 x 10$^{14}$cm$^{-2}$ [14,15]. The total amount of gold deposited during the high-temperature experiments was measured in monolayers as a product of the deposition rate and time. After the REM investigations, the samples were analyzed ex situ by atomic force microscopy (AFM) under ambient conditions.

2. Results and discussion

Fig. 1a shows an REM image of the central part of the crater that formed on the Si(111) surface after thermal annealing in an ultrahigh vacuum. The straight horizontal dark lines represent fragments of a circular-shaped atomic step a single atomic layer in height surrounding the central terrace approximately 100 μm in diameter. The corresponding AFM image of the large step-free region of the silicon is presented in Fig. 1d.

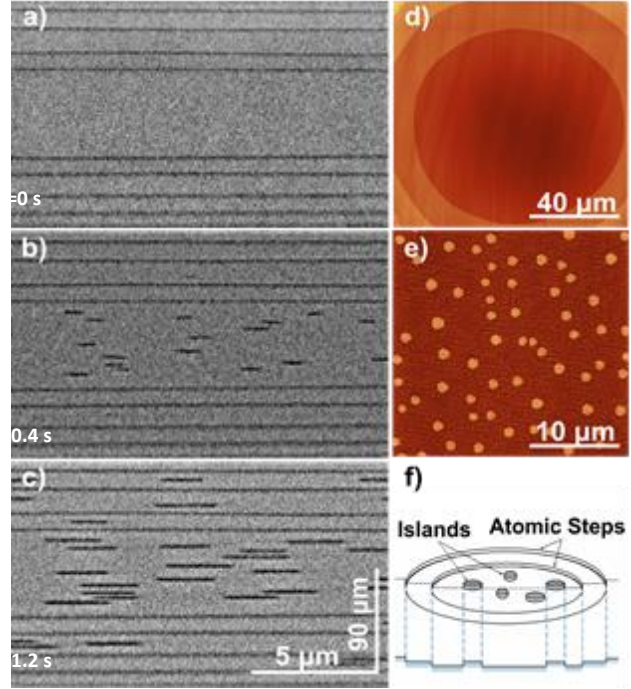

Fig. 1 REM (a)-(c) and AFM (d) and (e) images of the ultraflat Si(111) surface before (a) and (d) and after (b), (c), and (e) gold deposition at 930°C at the rate R = 0.031 ML/s. (f) Schematic representation of the surface structure.

During high-temperature annealing, atomic steps move in the direction of the higher situated terraces (outward from the center). In general, this motion results from the detachment of the atoms from the step edge subsequent to the surface diffusion of the adsorbed atoms on the terrace and evaporation in a vacuum. Because of this motion, the central terrace diameter increases over time. When the terrace size becomes larger than the critical value $D_c$, a new island nucleates near its center as a result of vacancy coagulation [16,17]. According to previously reported results, the critical diameter $D_c$ for island nucleation reaches as high as 120 μm at temperatures below 1000°C [12]. The period of island nucleation corresponds to the silicon sublimation rate, which is governed by activation energy of approximately 4.2 ± 0.2 eV. From this desorption barrier value, the time while the surface morphology of the central terrace remains unchangeable (until new island nucleation proceeds) can be estimated as approximately 27 s at 1100°C and more than 104 s at 900ºC. Therefore, the evaporation of the silicon atoms can be neglected at temperatures below 1000ºC, at least for the duration of experiments less than several tens of seconds.

However, the island nucleation time can be significantly reduced by the presence of residual oxygen in the atmosphere, which acts as an etchant of silicon [12,17,18]. As a result, the critical size of the terrace for the formation of new two-dimensional negative islands (etching pits) is smaller than that for ultrahigh vacuum conditions [12]. In our experiments, the size of the terrace at the bottom of the crater did not change significantly and remained below the critical value $D_c$ prior to gold deposition. No nucleation of new islands related to the thermal etching of the silicon surface by residual oxygen was detected.

The lack of stability on the Si(111) surface caused by the submonolayer gold deposition at 930ºC is illustrated in Fig. 1b and c. The appearance of the additional dark contrast features associated with new two-dimensional islands is visualized at the central terrace during the deposition of gold at the rate of 0.031 ML/s. The round two-dimensional islands are depicted as lines or strongly elongated ellipses in the REM images due to aforementioned foreshortening. During gold deposition, islands nucleate at the central terrace far from the other islands and the bounding step edge, as schematically shown in Fig.1f. Islands grow over time and cover almost all of the surface during gold deposition. At the same time, all of the steps in the field of view move in the direction corresponding to the capturing of adatoms in accordance with previously reported results [19].

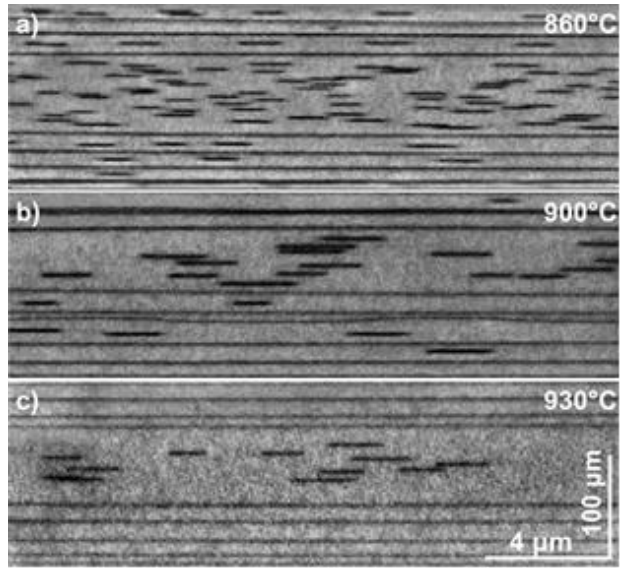

Fig. 2 REM images of the silicon terrace at the bottom of crater after the deposition of 0.04 ML of gold at 860ºC (a), 900ºC (b), and 930ºC (c).

As the substrate temperature increases, fewer islands nucleate and the distance between the islands increases as shown in Fig. 2. This behavior is very similar to the nucleation of two-dimensional islands during homoepitaxial growth, but in this case, no silicon is deposited at the surface, so the source of the atoms is not evident. If the steps are simply capturing gold atoms, the islands and step edges should consist of gold. Direct incorporation of the gold atoms into the step edges should result in stress formation along the atomic step, which generally can be visualized due to the diffraction contrast in REM [17,20]. However, our experiments revealed no differences in electron microscopy contrast from the steps on the atomically clean and gold deposited surface.

To explain the surface morphology transformations during high-temperature submonolayer gold deposition, the mass flow responsible for the growth of the islands cannot be supplied by the surrounding atomic steps but can be produced by the silicon bulk. To prove this, we first establish that the islands observed are the growth islands. Fig. 1e is a topographical AFM image of the central terrace with nucleated two-dimensional islands. All of the islands have the same height as a single atomic step. Moreover, the direct REM observation of the coagulation of the enlarging islands with descending steps during deposition also indicates that the islands and atomic steps have the same height. Considering that all of the atomic steps are moving in the growth direction during island nucleation, additional atoms cannot be supplied by the atomic steps. Therefore, the source of the atoms must be the silicon bulk.

Regarding the process of gold diffusion in silicon bulk, foreign atom diffusion in silicon can proceed via both direct and indirect mechanisms [6]. Indirect diffusion mechanisms of an impurity atom involve the formation of intrinsic point defects such as vacancies in the dissociative (Frank-Turnbull) mechanism [21] or self-interstitials in the so-called "kick-out" mechanism, which was first suggested for the diffusion of Al in Si by Watkins [22].

For gold diffusion in silicon, both the vacancy-mediated (Frank-Turnbull) and kick-out mechanisms are considered in the case of high density of the internal sinks or near the surfaces [23–26]. The interchange between the interstitial and substitutional sites in these mechanisms can modify the intrinsic point defect concentrations. In the Frank-Turnbull mechanism, the gold interstitials recombine with the vacancies, resulting in a decrease in the vacancy concentration in accordance with the quasi-chemical reaction $Au_i + V \leftrightarrows Au_s$, where $Au_i$, $Au_s$, and V represent the gold interstitial, gold substitutional, and vacancy, respectively. The lack of vacancies can be replenished by the diffusion from the silicon surface (the Schottky process) acting as sources for the vacancies. On the contrary, the kick-out mechanism is characterized by supersaturation of the self-interstitials, Aui ⇆ Aus + I. In this case, the excess of self-interstitials I is captured by the surface, which acts as an effective sink. Both processes (vacancy diffusion from the surface and annihilation of the self-interstitials at the surface) result in the net mass flux from the bulk to the surface.

2.1 Bulk diffusion model

To study the impact of the bulk processes on the surface morphology, the model illustrated in Fig. 3 is considered. The following notations are used. The gold is deposited at rate $R$ at the silicon surface at $x = 0$. The evaporation flux of the gold atoms is given by

$$R_e = g_i/\tau_{des},$$

where $g_i$ is the gold surface concentration, $\tau_{des}$ is the lifetime of the gold atoms prior to desorption, and $J_{gold}$ is the gold diffusion flux from the surface to the bulk. The bulk-surface exchange diffusion currents for silicon interstitials and vacancies are $J_I$ and $J_V$, respectively.

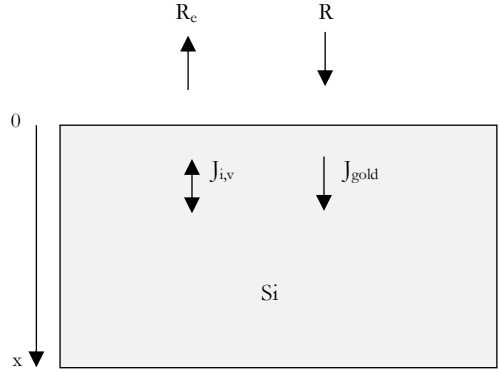

Fig. 3 Schematic representation of the net mass currents of the atoms and vacancies during the gold deposition onto the silicon terrace.

The full system of differential equations for the self-interstitials, vacancies, interstitial gold, and substitutional gold reported by Mathiot [26] were used:

$$\frac{\partial c_I}{\partial t} = D_I \frac{\partial^2 c_I}{\partial x^2} + k_{ko} c_i - k'_{ko} c_s c_I - k_{bm}(c_I c_V - c_I^* c_V^*) \tag{1}$$

$$\frac{\partial c_V}{\partial t} = D_V \frac{\partial^2 c_V}{\partial x^2} - k_{ft} c_V c_i + k'_{ft} c_s - k_{bm}(c_I c_V - c_I^* c_V^*) \tag{2}$$

$$\frac{\partial c_i}{\partial t} = D_i \frac{\partial^2 c_i}{\partial x^2} - k_{ft} c_V c_i + k'_{ft} c_s - k_{ko} c_i + k'_{ko} c_s c_I \tag{3}$$

$$\frac{\partial c_s}{\partial t} = k_{ft} c_V c_i - k'_{ft} c_s + k_{ko} c_i - k'_{ko} c_s c_I. \tag{4}$$

Here, $D_I$, $D_V$, and $D_i$ and $c_I$, $c_V$, and $c_i$ denote the diffusion coefficients and concentrations of the self-interstitials, vacancies, and gold interstitials, respectively, and $c_s$ is the substitutional gold concentration. The corresponding equilibrium concentrations of the point defects are indicated by asterisks. The first terms in equations (1)-(3) represent diffusion caused by differences in the concentrations of the species. It is assumed that the gold atoms in the substitutional sites are totally immobile, so the diffusion term is canceled in equation (4). The last term in equations (1) and (2) describes the processes of the bulk creation and annihilation of the self-interstitials $I$ and vacancies $V$ (Frenkel pairs).

The rate constants for the forward $k_{ko}$ and $k_{ft}$ and backward $k'_{ko}$ and $k'_{ft}$ kick-out and Frank-Turnbull reactions are given by [26]:

$$k_{ko} = 4\pi r_{ko} D_i N_s, \tag{5}$$

$$k'_{ko} = 4\pi r'_{ko} D_I, \tag{6}$$

$$k_{ft} = 4\pi r_{ft}(D_i + D_V), \tag{7}$$

$$k'_{ft} = v_0 exp(-E'_{ft}/kT), \tag{8}$$

where $N_s = 5.0 \times 10^{22}$ cm$^{-3}$ is the lattice site density and $r_{ko}$, $r'_{ko}$, and $r_{ft}$ are the effective capture radii for the forward kick-out, backward kick-out, and forward Frank-Turnbull reactions, respectively. Equation (8) describes the backward reaction for the vacancy mechanism, where $v_0 = 2.59 \times 10^{13}$ s$^{-1}$ is an attempt frequency [27], $E'_{ft}$ is the activation barrier, $k$ is the Boltzmann constant, and T is the absolute temperature. The capture radii $r_m$, where subscript $m$ refers to the forward kick-out (*ko*), forward Frank-Turnbull (*ft*), and backward kick-out (*ko'*) reactions, are on the order of a few angstroms and can be described by including the Boltzmann exponent with the corresponding energy barriers ($E_m$):

$$r_m = r_m^* exp(-E_m/kT)$$

The initial concentrations of the substitutional $c_s$ and interstitial $c_i$ gold are assumed to be zero everywhere in the sample at t = 0. Consequently, the initial concentrations of the vacancies $c_V$ and self-interstitials $c_I$ have thermal equilibrium values of $c_V^* = 5.0 \times 10^{22} e^{3.15} e^{-(3.36\ eV/kT)}$ cm$^{-3}$ and $c_I^* = 2.9 \times 10^{24} e^{-(3.18\ eV/kT)}$ cm$^{-3}$.[26] The diffusion coefficients for the self-interstitials and vacancies $D_I = N_s \times 819^{+407}_{-309} \times e^{-((4.82\pm0.05)\ eV/kT)}/c_I^*\ cm^2 s^{-1}$ and $D_V = N_s \times 206^{+112}_{-72} \times e^{-((4.65\pm0.05)\ eV/kT)}/c_I^*\ cm^2 s^{-1}$ are deduced from self-diffusion experiments that were performed at temperatures between 750°C and 1400°C [5]. The temperature dependence of $D_i = 0.28 \times$ $e^{-(1.6\,eV/kT)} cm^2 s^{-1}$ follows from the analysis of the diffusion of ion-implanted gold in a silicon bulk [28].

Since we analyzed the initial stages of the metal diffusion from the silicon surface, it is thus essential to carefully choose the boundary conditions. Under common experimental conditions, a silicon wafer is usually covered with thick layers of metal acting as an infinite source of foreign atoms. Therefore, diffusion equations are usually solved using Dirichlet-type boundary conditions: $c_s = c^*_{Au}$ on the gold-covered surface and accumulation or evaporation of the impurity at the back side of the wafer (Neumann-type boundary conditions). Assuming local equilibrium at the surface boundary between the gold layer and the silicon, the gold interstitial concentration takes the form $c_i = f_i c^*_{Au}$, where $c^*_{Au}$ is the total gold solubility and $f_i$ is the fraction of the interstitial gold under equilibrium conditions [29]. These assumptions are validated in long-duration diffusion experiments showing that the equilibrium concentration of gold is preserved at the surface during annealing [30,31]. However, at the initial deposition stage when local equilibrium is not attained, Dirichlet-type boundary conditions are not suitable because the equilibrium concentration of gold is not established at the surface.

If gold is deposited on an atomically clean surface, the boundary condition for Au$_i$ is formulated in terms of the limitation of the transition flux through the surface, that is:

$$-D_i \frac{\partial c_i}{\partial x}\bigg|_{x=0} = \frac{g_i}{\tau_i} - k^i c_i(x=0, t), \tag{9}$$

where $k^i = 1/[e^{(E_i/kT)} - 1]$ is the kinetic coefficient for the gold atom penetration through the silicon surface and $E_i$ is the energy barrier for the gold atom penetration, which can vary from zero (complete permeability) to infinity (complete impermeability). The first term in equation (9) corresponds to the flux of gold into the bulk $J_{gold} = g_i/\tau_i$, where $g_i$ is the actual surface concentration of the gold atoms and $\tau_i$ is the lifetime of the gold atoms prior to dissolving in bulk. Neumann boundary conditions are used on the other side for interstitial gold.

At high temperatures, the sublimation of the gold atoms from the surface must be considered. The gold surface concentration $g_i(t)$ is given by:

$$\frac{\partial g_i}{\partial t} = R - g_i/\tau, \tag{10}$$

where $\tau$ is the mean lifetime of an adsorbed gold atom and $R$ is the deposition rate. The second term in equation (10) is the current of gold atoms leaving the surface, which actually consists of the sum of two fluxes $J_{gold} = g_i/\tau_i$ that describe the process of dissolving in bulk and $R_e = $ $g_i/\tau_{des}$ that is due to the evaporation of the gold atoms. The total lifetime is then $\frac{1}{\tau} = \frac{1}{\tau_{des}} + \frac{1}{\tau_i}$, where $\tau_{des} = \nu_0^{-1} e^{E_{Au}/kT}$. Thus, the concentration of the deposited gold atoms is given by:

$$g_i = R\tau \left(1 - e^{-t/\tau}\right), \quad (11)$$

where $t$ is the deposition time. Experimentally determined activation energies for gold atom desorption from a silicon surface have been reported in the range of 3.3-3.7 eV [32]. These results are considering that gold losses due to desorption can be described by the mean lifetime of adsorbed gold atom $\tau_{des}$ =10-560 s at 900ºC. Since the measurement time did not exceed 10 s in our experiments, the effect of gold evaporation was also neglected at 900ºC. However, at higher temperatures, gold desorption must be considered for the correct analysis of the experimental results.

The boundary conditions for the self-interstitials and vacancies are formulated considering the limited efficiency of the annihilation and the creation of point defects at the surface [33]:

$$-D_m \frac{\partial c_m}{\partial x}\Big|_{x=0} = k^m[c_m^* - c_m(x=0,t)], \quad (12)$$

where the subscript $m$ represents either $I$ or $V$ and $D_m$ and $c_m$ are the diffusion coefficient and concentration of a corresponding point defect, respectively. The same type of equation (with the reverse sign on the right-hand side) is chosen at x = L at the back side of the sample. The kinetic coefficient $k^m = 1/[e^{(E_m/kT)} - 1]$, where $E_m$ is the activation energy for the interaction of the corresponding point defect with the surface. In our simulations, $E_V$ is the order of the binding energy of the vacancies to the clusters 3.2 ± 0.2 eV [34]) and $E_I$ is the simulation parameter.

The system of equations (1)-(4) was solved numerically using COMSOL commercial software. The results of the simulations were verified using a self-written Fortran code, where the backward time-centered space discretization scheme was implemented. Fig. 4 represents the kinetic behavior of $c_I$ and $c_V$ in the bulk silicon predicted by solving equations (1)-(4). The sample thickness L is equal to 300 μm, the gold deposition rate is 0.022 ML/sec, and the sample temperature is 900ºC. An increased self-interstitial concentration due to gold in-diffusion is found in the near-surface area. At the same time, the vacancies are undersaturated because of recombination with the self-interstitials. However, an increase in $c_V$ is observed in the subsurface area. This indicates that a surface acting as a natural source of vacancies produces the vacancy flux. Similar behavior of the vacancy concentration was found in the numerical analyses of the coupled equations for the interstitial and vacancy diffusion in silicon in prior studies [3,35].

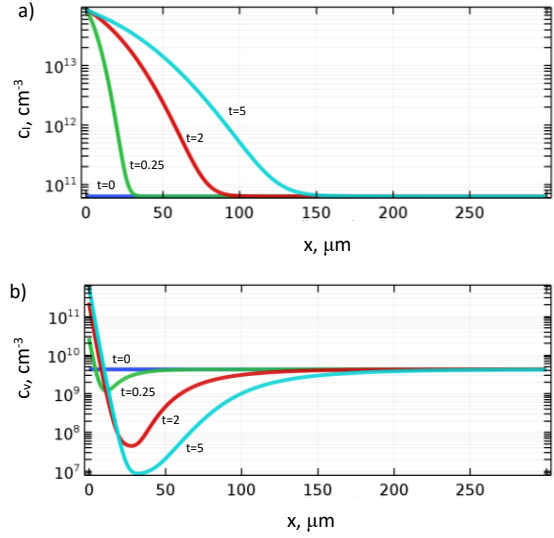

Fig. 4 Silicon self-interstitial (a) and vacancy (b) concentration depth profiles. The profiles represent solutions of equations (1)-(4) for different times t = 0, 0.25, 2, and 5 s. The sample temperature is 900°C and the gold deposition rate R = 0.022 ML/s.

Another source for vacancies is the backward Frank-Turnbull reaction. According to this dissociative reaction, the gold atoms dissolved in the substitutional site move in the interstitial position, leaving bulk vacancies.

On the basis of the calculated concentration profiles of the intrinsic defects, we analyzed how the changing of the point defect concentration in the silicon bulk can affect the surface processes. The out-diffusion of the self-interstitials and the in-diffusion of the vacancies both produce the positive flux of the material being transported from the bulk to the surface. The gradients of the concentrations normal to the surface produce a net mass flux:

$$J_m = -D_m \frac{\partial C_m}{\partial x}|_{x=0}$$

out (or into) the bulk of the silicon to restore the equilibrium concentrations of the intrinsic bulk point defects. Herein, the subscript $m$ refers to the interstitial $I$ or vacancy $V$. The total flux of the material added to the surface is determined by adding the interstitial flux out of the bulk and the vacancy flux into the bulk:

$$J = J_I + J_V. \tag{13}$$

The vacancy flux $J_V$ is negligibly small compared with $J_I$ and thus the Frank-Turnbull mechanism is not significant and does not alter the total flux. However, the concentration profiles depend on whether they are calculated considering only the kick-out mechanism or both mechanisms. Since the silicon interstitials are the major contributors to the flux, for simplicity, the impact of the vacancy-surface interactions will be neglected below.

2.2 Surface kinetics

The nucleation and growth of the two-dimensional islands provide a substantial understanding of the surface mass transport in molecular beam epitaxy [29,36], equilibrium [37], oxygen etching [16,17,38], and sublimation [39]. Detailed reviews of the nucleation and growth processes can be found in the literature [29,36,40]. To explain the surface morphology transformations, we analyzed the growth kinetics of the two-dimensional islands that formed during high-temperature gold deposition. Video sequences of the recorded REM images (similar to those presented in Fig. 1 and Fig. 2) were analyzed frame by frame and the island sizes were measured at 860-1000°C and the gold deposition rates in the range of 0.01-0.05 ML/s (Fig. 5). The area of each island was calculated assuming the circular shape of the islands. This assumption is reasonable at sample temperatures above 830°C where no surface reconstruction exists on an atomically clean Si(111) surface. Furthermore, the surface diffusion and step line tension are expected to be nearly anisotropic at these temperatures [27]. Direct AFM measurements of the quenched samples after submonolayer gold deposition at high temperatures confirmed the round island shape (Fig. 1e). Therefore, we treated the islands as circular in shape.

To investigate the dependence of the growth rates on the local environment, the areas of the surface closest to each island were analyzed (Fig. 5). Fig. 5b shows the time dependence of the area for the individual islands represented via an REM image (Fig. 5a). All of the islands grow linearly in time but the growth rates are different. The islands with smaller capture zones and closer situated neighboring islands (for example, islands 5 and 6) grow slower than the islands nucleated apart from the other islands and the atomic steps (for example, islands 4 and 2). The maximum and minimum of the growth rates are measured for Island 4 and Island 6, respectively. Fig. 5c represents the temporal dependences of the total area covered by the islands at different sample temperatures. The linear kinetics of the island area growth were preserved up to the 1000°C sample temperature.

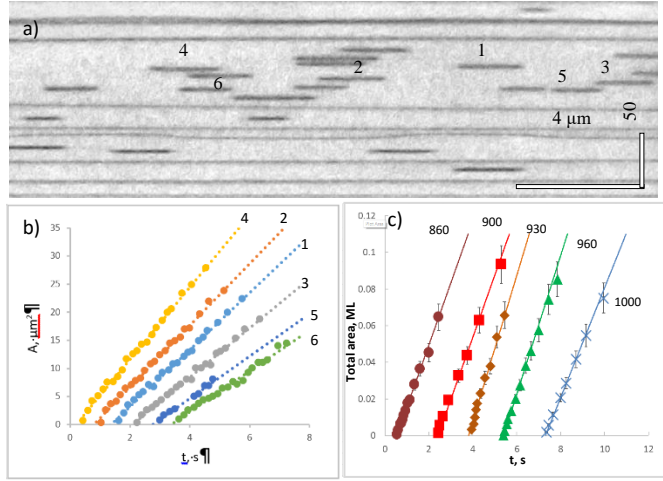

Fig. 5. (a) REM image of the islands at T = 900°C. The analyzed islands are labelled by numbers (1)-(6). (b) Time dependence of the individual island areas (1)-(6). The deposition rate is 0.021 ML/s and the measured growth rates are (1) 5.0, (2) 5.5, (3) 4.4, (4) 6.4, (5) 3.8, and (6) 3.6 µm²/s. (c) Experimental time dependences of the total island area at different sample temperatures. The deposition rates are: 0.032 (860°C), 0.031 (900°C), 0.041 (930°C), 0.035 (960°C), and 0.028 ± 0.007 ML/s (1000°C). The experimental data are represented by the symbols and the solid lines show the calculated area dependences. The curves in (b) and (c) are shifted in time for clarity.

Considering the growth of two-dimensional islands during gold deposition, we followed the approach used by Stoyanov and Michailov [41][45] and Markov [36] and later elaborated by Herviue et al. [42]. In our model, the net mass flux from the silicon bulk produces increased concentrations of silicon atoms at the atomically flat surface during gold in-diffusion. The increase in the silicon adatom concentration results in high supersaturation and therefore the formation of new two-dimensional islands.

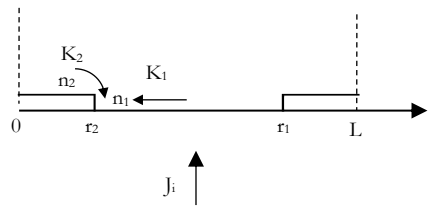

Fig. 6 The surface model for the calculation of the island growth rate.

The nucleation of the island occurs when the two adsorbed species meet. The growth of the nucleated islands proceeds via capturing randomly moving atoms. In the calculations to follow, it is assumed that the nucleated islands are regularly distributed over the surface area. The distance between the island centers is L and all of the islands are the same size (Fig. 6). The positions of the step edges surrounding the neighboring islands are denoted by $r_1$ and $r_2$. The center of the left island is considered the origin. The adatom concentration $n(r)$ on the surface can be found by solving the diffusion problem in the polar coordinates, which in the case of complete condensation is:

$$D_s \frac{d^2 n}{dr^2} + \frac{1}{r}\frac{dn}{dr} + J = 0. \tag{14}$$

Here, $n(r)$ is the surface concentration of the silicon adatoms, $D_s$ is the diffusion constant, $J$ is the net mass flux deduced from equation (13), and $r$ is the coordinate. The solution for the adatom concentration on the terrace between the islands ($r_2 < r < r_1$) is given by [36]

$$n_1(r) = \widetilde{n_1} + \frac{J}{4D_s}(r_1^2 - r^2) - \frac{J}{4D_s}\frac{(r_1^2 - r_2^2)}{Ln\left(\frac{r_2}{r_1}\right)} Ln\left(\frac{r}{r_1}\right),$$

where $\widetilde{n_1}$ is the adatom concentration in the vicinity of the descending step. The adatom concentration at the top of the island ($0 < r < r_2$) is

$$n_2(r) = \widetilde{n_2} + \frac{J}{4D_s}(r_2^2 - r^2),$$

where $\widetilde{n_2}$ is the concentration of silicon adatoms in the vicinity of the ascending step. The concentrations $\widetilde{n_1}$ and $\widetilde{n_2}$ can be found from the boundary conditions:

$$-D_s \frac{dn_1}{dr}\bigg|_{r=r_2} = -K_1\left(\widetilde{n_1} - n_{eq}(r_2)\right),$$

$$-D_s \frac{dn_2}{dr}\bigg|_{r=r_2} = +K_2\left(\widetilde{n_2} - n_{eq}(r_2)\right),$$

where $n_{eq}(r)$ is the equilibrium adatom concentration and $K_1$ and $K_2$ are the kinetic coefficients for the attachment to the step that encloses the island from the top terrace and the region around the island, respectively.

According to the models of island growth and decay previously reported, the island kinetics are governed by the net adatom current density at the perimeter of the island [27,39]. The net diffusion flux densities of the adatoms to the island edge arriving from the lower and upper terraces are given by:

$$j_1 = -D_s \frac{dn_1}{dr}\bigg|_{r=r_2} = \frac{Jr_2}{2}\left(1 + \frac{(r_1^2 - r_2^2)}{2r_2^2 Ln\left(\frac{r_2}{r_1}\right)}\right),$$

$$j_2 = -D_s \frac{dn_2}{dr}\Big|_{r=r_2} = \frac{Jr_2}{2}.$$

The rate of the island growth is given by the sum of the fluxes:

$$V = \frac{dr_2}{dt} = \Omega(j_1 + j_2) = \frac{\Omega J}{4} \frac{(r_1^2 - r_2^2)}{r_2 Ln\left(\frac{r_1}{r_2}\right)},$$

where $\Omega$ is the atomic surface area.

$$V = \frac{\Omega J}{2\pi N_s r_2} \frac{(1-2\varphi)}{Ln\left(\frac{(1-\varphi)^2}{\varphi^2}\right)}, \tag{15}$$

where $\varphi = {r_2}/{L}$ and $N_s$ is the island surface density. The rate of change of the island area is

$$\frac{dA}{dt} = \Omega(j_1 + j_2) * 2\pi r_2 = \frac{\Omega J}{N_s} \frac{(1-2\varphi)}{Ln\left(\frac{(1-\varphi)^2}{\varphi^2}\right)} \tag{16}$$

When $r_2 \ll L$, equation (16) simplifies to

$$\frac{dA}{dt} = \frac{\Omega J}{N_s} \tag{17}$$

This is the same result obtained for the case of complete condensation and impermeable steps [36,42]. If the density of the nucleated islands remains unchangeable and the flux $J = const$, the island area growth rate is constant and this gives the experimentally observed linear dependence of the island area on the time. In our experiments, the spacing between the islands is measured in a range of 6.8 to 10.1 μm. Substituting flux $J$ calculated from the concentration profiles of the intrinsic defects, the respective growth rates are 3.0 and 6.6 μm²/s. The latter agrees with the experimentally measured island kinetics (Fig. 5).

As it is not easy to investigate the lateral distribution of the islands in REM due to the aforementioned foreshortening, we analyzed the total area covered by the islands (Fig. 5c). The integration of equation (17) gives $AN_s = \Omega J t$, where the left part is the fraction of the surface covered by the islands. The best fit of the experimental data in a temperature range of 860-1000°C represented in Fig. 5c by solid lines is obtained with $E_i = 2.7 \pm 0.2$ eV and $E_I = 1.5 \pm 0.2$ eV. The deposition rates used in the model are 0.034 (860°C), 0.034 (900°C), 0.040 (930°C), 0.038 (960°C), and 0.030 ML/s (1000°C), which are very close to the experimentally measured error. The obtained value of the energy barrier $E_i$ agrees well with the first principle evaluation of the penetration energy of gold into the silicon substrate [43].

It is assumed that the self-interstitials diffuse to the surface until they attach to the step surrounding the terrace or island edge and eventually incorporate into the step. To analyze the total amount of the material coming from the crystal bulk, we consider the case of complete condensation [36]. The evaporation process can be neglected in this nucleation regime. This means that all of the interstitials eventually join the growing islands or atomic steps.

To justify the implementation of the complete condensation theory, the evaporation rate must be considered with the net current of the additional atoms to the surface. According to the literature, the data silicon sublimation rate can be described via Arrhenius dependence with activation energy of approximately $4.2 \pm 0.2$ eV [44]. The corresponding evaporation rate can be estimated as $5 \times 10^{-6}$ ML/s [44–46]. This means that during the time of measurements (several seconds), only a small fracture of the adatom can be evaporated from the surface at 900°C. However, at higher temperatures, the evaporation rate increases. For example, at 1100°C, one monolayer of silicon evaporates every 27 s. Thus, the finite adatom lifetime must be considered and the complete condensation case is not applicable at high sample temperatures.

According to theory, surface morphology can be affected by the kinetic limitations during two-dimensional island growth or decay [27,47–49]. The rate-limiting kinetics can be characterized as diffusion-limited (DL) or attachment-detachment limited (ADL) depending on the relative importance of the surface diffusion or the attachment-detachment of the adatoms at the step edge, respectively [50]. Earlier it was shown that the step motion on the Si(111) surface is described by the DL-kinetic regime rather than the ADL conditions [27,39,51]. On the contrary, atomic step kinetics and two-dimensional negative island nucleation on an Si(111) surface in the case of a high concentration of surface vacancies is described by the attachment-detachment kinetics, with the activation energy of the vacancy-step interaction larger than the surface diffusion barrier [12,16,52].

The dependence of the island growth rates on the distance to the neighboring islands suggests that island kinetics are controlled by the rate of the attachment and detachment process at the island edges [17,53]. If the energy barrier for adatom attachment to the step edge is small enough with respect to the diffusion barrier, there is a gradient of the chemical potential across the terrace that is determined by the terrace width and the difference in the step chemical potentials. In the opposite case when the adatom diffusion is fast and the rate limiting factor is the attachment-detachment process, the chemical potential is constant along the terrace. The island growth rate is proportional to the differences between the chemical potentials at the step and adjacent terrace. Thus, the linear dependence of the island growth or decay provides evidence of the attachment-detachment limited atomic mechanism of the surface transport. The latter is consistent with the experimental observation of Ostwald ripening [37] and thermal etching [17] of the Si(100) surface.

The nonlinear kinetics of the island growth or decay usually observed in silicon homoepitaxy experiments are the consequence of the diffusion-limited (DL) mechanism of the surface transport. In this case, the nucleated islands grow as $S \sim t^{2/3}$. For an Si(111)-(1 x 1) surface, the experimental observations of the individual island decay process and the atomic step fluctuations revealed the diffusion-limited and not the attachment-detachment limited mechanism of the surface transport [27,54,55]. The decay kinetics of the isolated islands and holes was described by power law dependence with an exponent value of 0.78 [54]. The evidence of the diffusion-limited mechanism of the surface mass transport was found in studies of surface morphology transformations during sublimation [56,57] and epitaxial growth [58].

The classic crystal growth theory predicts that the growth of two-dimensional islands proceeds via the formation of stable nuclei of size $i$ (where $i$ is the number of atoms) greater than critical size $i^*$ [36]. The nucleus density that is equal to the squared island spacing $L^2$ depends on the deposition rate $R$ and substrate temperature $T$:

$$\frac{1}{N_s} = L^2 \sim R^{-\chi} e^{-\frac{E^*}{kT}}, \qquad (18)$$

where $E^*$ is the energy gain when the critical nucleus is formed, $\chi$ is the scaling exponent, and k is Boltzmann's constant.

Fig. 7 shows the experimentally measured dependence of the inverse island density $N_s^{-1}$ on the calculated interstitial flux $J$ at 900°C. After the nucleation of the islands at the silicon terrace, the sample was quenched to room temperature. The sample cooling rate was estimated to be 400°C/s. Subsequent measurements of the island spacing were done using AFM. To avoid the impact of atomic steps, the average distance between the islands is measured only in the central part of the terrace. The deposition rate $R$ in equation (18) is equivalent to the interstitial flux $J$ calculated on the basis of the concertation profiles (Fig. 4).

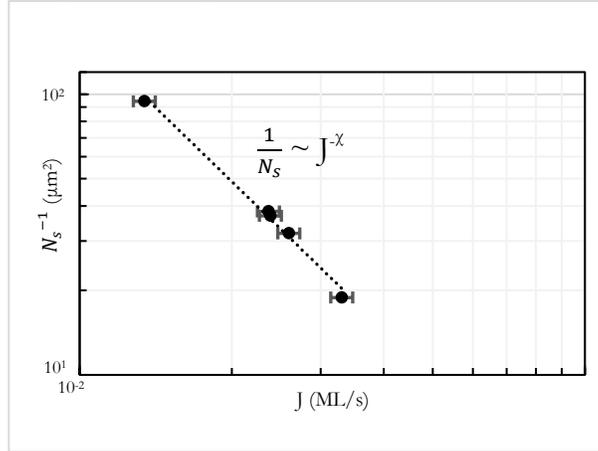

Fig. 7. Experimental dependence of $N_s^{-1}$ on the calculated deposition rate $J$ at 900°C.

By extracting a slope from the $N_s^{-1}(J)$ measurements taken over a range of fluxes, we extract the scaling exponent $\chi$ = 1.72-1.76. According to the theoretical approach, parameter $\chi$ appears to be below 1 for any $i > 0$ under DL conditions [59]. In the case of attachment-detachment limited growth (ADL) conditions, parameter $\chi$ is greater than 1. Thus, the experimentally measured value of $\chi$ is attributed to the attachment-detachment regime and corresponds to the critical nucleus size $i^* = $ 19-22 via the formula $\chi = 2i^*/(i^* + 3)$.

The obtained results can be compared with the data from homoepitaxial growth experiments. Earlier it was shown that the two-dimensional island growth kinetics at the Si(111) surface are characterized by the diffusion-limited regime at a temperature range of 720-1090°C [60,61]. The critical nucleus size $i^*$ was estimated in the range of 1-12 atoms at a low temperature range of 720-760°C. Recent electron microscopy studies of homoepitaxial growth revealed the increase in $i^*$ and $\chi$ at elevated temperatures $i = 16 \pm 11$, $\chi = 0.82 \pm 0.11$ (950°C)[64] and $i = 25 \pm 13$, $\chi = 0.9 \pm 0.05$ at 900-1180°C. The values of $\chi$ correspond to the diffusion-limited kinetics of two-dimensional island growth.

Our experimentally obtained value of the critical nucleus size (19-22 atoms) agrees within the error limits of these previously reported values. This fact suggests the same mechanism for island nucleation at a silicon terrace during gold deposition and high-temperature homoepitaxial growth. The nucleation process can be briefly described as follows: the diffusion of gold atoms from the surface to the bulk produces an excess of self-interstitials in the subsurface area via the kick-out mechanism. The net mass current of the nonequilibrium self-interstitials to the surface results in the increased concentration of adatoms due to the penetration of the self-interstitials through the surface boundary and therefore the formation of new two-dimensional islands.

The linear time dependence of the island sizes and scaling exponent value χ = 1.72-1.76 indicate that island growth is limited by the attachment-detachment kinetics in the case of gold deposition. The relative importance of the surface diffusion and the step attachment rate is characterized by kinetic length $d = D/K$, where $D$ is the diffusion constant and $K$ is the attachment-detachment kinetic coefficient [55,62]. The diffusion-limited and attachment-detachment extremes correspond to $d = 0$ and $d = \infty$, respectively. This approach also accounts for asymmetry in the attachment-detachment kinetics at the atomic step (the Ehrlich-Schwoebel (ES) barrier) [54]. If there are no ES barriers (the instantaneous attachment of the adatoms to the step), $d = 0$ is a pure diffusion-limited regime. The increase in the attachment-detachment barrier at the step produces an increase in the kinetic length that in turn induces the change in the kinetic regime from diffusion-limited to attachment-detachment limited.

The adatom-step interaction can be affected by the presence of foreign atoms at the silicon surface that act as surfactants [63]. Prior studies established that silicon deposition on an Si(111) surface covered by superstructural reconstruction (5 x 2) initiated by gold adsorption results in a lower density of 2D islands than on an atomically clean surface [64]. The depression of two-dimensional island nucleation can be understood in terms of the increase in the diffusion length of the silicon and the change in the kinetics of atom incorporation into the step at the gold deposited surface. The increase in the diffusion coefficient can be addressed by the structural transformations of the first atomic layer due to the formation of surface reconstructions. However, in our experiments, no RHEED patterns attributed to the formation of new superstructural reconstructions were observed during gold deposition at temperatures above 900ºC in accordance with previously reported studies. This indicates that changes in the diffusion length can probably be neglected in treatments of island nucleation.

When the adatoms attach to the steps on the crystal surface, they diffuse along the step edge and incorporate in the kink sites. The activation energy of the adatom-step interaction is related to the step kink density [65]. When the kink density is high, adatoms easily attach to the step. However, when the kink density is low, it is possible for adatoms to detach from the step without solidification. Thus, smoothing the step edges that correspond to lowering the kink density may increase the energy barrier for adatom-step interactions. The increased energy barrier for silicon adatom attachment to the step may explain the attachment-detachment limited conditions during gold deposition.

Silicon can form eutectic alloys with gold. The eutectic point is approximately 359°C and well below the respective melting points of gold (1063°C) and silicon (1412°C) [66]. One important question is how the eutectic reaction proceeds at the beginning of the process of gold deposition onto a silicon surface. The birth stages of eutectic formation were investigated by field-ion microscopy at 450°C [67]. The submonolayer deposition of gold on a Si(111) surface results in the formation of an intermixed layer with small clusters consisting of several gold atoms. When the amount of deposited gold increased to 2 ML, the eutectic reaction produced gold rods in the silicon phase. These findings agree with investigations of three-dimensional island nucleation at an Si(111) surface during high-temperature gold deposition [14]. Three-dimensional island nucleation starts at coverages above 0.7 ML for sample temperatures in a range of RT-750°C. At temperatures above the eutectic temperature, the three-dimensional islands consist of a mix of gold and silicon atoms [In]. Recent studies showed that the incorporation of silicon atoms could be detected by analyzing the Si-Au 3D island motion at the Si(111) surface [68,69]. In our experiments, no 3D island formation was observed during gold deposition at gold coverages well below 0.7 ML. Thus, there is no evidence of eutectic formation at the silicon surface in our experimental conditions.

The formation of an eutectic layer at the silicon surface when the gold coverage exceeds 0.7 ML have implications for analyzing the gold bulk diffusion in silicon. Obviously, the equilibrium gold concentration in the gold-silicon alloy is preserved at given temperatures (as defined by the phase diagram). Thus, the boundary conditions in equation (9) can be replaced by Dirichlet-type with the respective equilibrium concentration for $Au_i$. According to our calculations, this change in the boundary conditions will result in a symmetric U-shaped concentration profile of gold diffusing from the surface as previously reported [70–72]. This means that Dirichlet-type boundary conditions for $Au_i$ are useful for the correct simulation of long-duration gold diffusion in silicon, whereas initial depth profiles of native point defects and impurity concentrations form at the start of metal diffusion and are determined by equation (9).

3. Conclusion

In summary, this study employed in situ reflection electron microscopy in conjunction with the developed method of large area step-free Si(111) terrace formation to study in real time the interaction of self-interstitials with the crystal surface. The submonolayer gold deposition onto the step-free Si(111) surface at high temperatures (860-1000°C) is supplemented by the self-interstitial formation in the subsurface area and the consequent out-diffusion to the surface sinks, which initiates the formation of two-dimensional islands. By solving the diffusion equations for bulk and surface diffusion, the interstitial flux $J$ to the surface is found to be restricted by the activation barrier of $1.5 \pm 0.2$ eV, which corresponds to self-interstitial penetration through a silicon terrace. The process of gold dissolution from a silicon surface is characterized by the activation energy $2.7 \pm 0.2$ eV. The kinetic growth behavior of the nucleated two-dimensional islands is dominated by the reactions at the atomic step edges. The obtained results also reveal that the diffusion of gold atoms in silicon proceeds via the kick-out mechanism even in the presence of sinks and sources of point defects. The results of this study provide a comprehensive description of point defect diffusion and the interaction with the surface sinks, which should have important implications not only for understanding the atomic mechanisms of nanoscale object formation, but also for the modelling of such processes as supercooling, surface melting, surface eutectic formation, nanowires, and surfactant-mediated growth.

4. Acknowledgments

S.S., L.F, and A.L gratefully acknowledge partial financial support for the in situ REM experimental investigations from the Russian Science Foundation (grant 14-22-00143).